\documentclass{article}
\usepackage{spconf}
\ninept 

\usepackage[utf8]{inputenc} 
\usepackage[T1]{fontenc}    

\usepackage{url}            
\usepackage{booktabs}       
\usepackage{amsfonts}       
\usepackage{nicefrac}       
\usepackage{microtype}      
\usepackage{amsmath}        
\usepackage{amssymb}
\usepackage{amsfonts}
\usepackage{amssymb}
\usepackage{siunitx}
\usepackage{float}
\usepackage{caption}
\usepackage{enumitem}
\usepackage{lipsum}
\usepackage{algorithm}
\usepackage{algorithmic}
\usepackage{graphicx}
\usepackage{textcomp}
\usepackage{xcolor}
\usepackage{mathrsfs}
\usepackage{bm}

\usepackage{hyperref}       
\hypersetup{allcolors=blue,colorlinks=true}

\usepackage{mathtools}
\usepackage{tcolorbox}
\usepackage{color}
\usepackage{theorem}
\usepackage{caption}
\usepackage{subcaption}

\newcommand{\R}{\mathbb{R}}
\newcommand{\bw}{\bm{w}}
\newcommand{\bx}{\bm{x}}
\newcommand{\bz}{\bm{z}}
\newcommand{\bD}{\bm{D}}
\newcommand{\bI}{\bm{I}}

\newcommand{\bM}{\bm{M}}
\newcommand{\bU}{\bm{U}}
\newcommand{\bX}{\bm{X}}
\newcommand{\Gcal}{\mathcal{G}}
\newcommand{\Xcal}{\mathcal{X}}
\newcommand{\Ucal}{\mathcal{U}}
\newcommand{\Vcal}{\mathcal{V}}
\newcommand{\Zcal}{\mathcal{Z}}
\DeclareMathOperator*{\argmin}{arg\,min}
\DeclareMathOperator{\prox}{\textsf{prox}}

\newtcolorbox{myblockt}[1]{colback=urblue!5!white,
	colframe=urblue,fonttitle=\bfseries,
	title=#1}

\newtcolorbox{myblock}{colback=urblue!5!white,
	colframe=urblue,fonttitle=\bfseries}

\def\BibTeX{{\rm B\kern-.05em{\sc i\kern-.025em b}\kern-.08em
    T\kern-.1667em\lower.7ex\hbox{E}\kern-.125emX}}

\newcommand{\email}[1]{\href{mailto:#1}{#1}}
\newcommand\mad[1]{\noindent{\textcolor{blue}{[Mad: #1]}}}
\ninept

\makeatletter
\def\twoauthors#1#2#3#4{\gdef\@address{}
   \gdef\@name{\begin{tabular}{@{}c@{}}
        {\em #1} \\ \\
        #2\relax
   \end{tabular}\hskip 0.5in\begin{tabular}{@{}c@{}}
        {\em #3} \\ \\
        #4\relax
\end{tabular}}}
\makeatother

\title{Network Clustering for Latent State and Changepoint Detection}

\twoauthors{Madeline Navarro and Genevera I. Allen\sthanks{GIA is also affiliated with the Departments of Statistics and of Computer Science at Rice University and with the Neurological Research Institute at Baylor
College of Medicine. GIA acknowledges support from NSF DMS-1554821, NSF NeuroNex-1707400,
and NIH 1R01GM140468}}%
             {Department of Electrical and Computer Engineering \\
              Rice University\\
              Houston, TX 77005 \\
              \email{nav@rice.edu} and \email{gallen@rice.edu}}
              {Michael Weylandt\sthanks{MW's research is supported by an appointment to the Intelligence Community Postdoctoral Research Fellowship Program at the University of Florida Informatics Institute, administered by Oak Ridge Institute for Science and Education through an interagency agreement between the U.S. Department of Energy and the Office of the Director of National Intelligence.}}%
            {University of Florida Informatics Institute \\
             University of Florida \\
             Gainesville, FL 32611 \\
             \email{michael.weylandt@ufl.edu}}%

\usepackage[natbib=true,
            bibstyle=ieee,
            citestyle=numeric-comp,
            uniquename=false,
            uniquelist=false,
            maxcitenames=2,
            mincitenames=1,
            maxbibnames=6,
            mincrossrefs=99]{biblatex}
\renewbibmacro{in:}{\ifentrytype{article}{}{\printtext{\bibstring{in}\intitlepunct}}}
\DefineBibliographyStrings{english}{url = Available at}

\AtEveryBibitem{%
  \clearlist{publisher}%
  \clearfield{pagetotal}%
  \clearfield{series}%
  \clearfield{edition}%
  \clearfield{isbn}%
  \clearname{editor}%
  \clearlist{location}%
  \clearfield{url}
}

\begin{document}
\begin{refsection}

\maketitle

\begin{abstract}
Network models provide a powerful and flexible framework for analyzing a wide range of structured data sources. In many situations of interest, however, multiple networks can be constructed to capture different aspects of an underlying phenomenon or to capture changing behavior over time. In such settings, it is often useful to \emph{cluster} together related networks in attempt to identify patterns of common structure. In this paper, we propose a convex approach for the task of network clustering. Our approach uses a convex fusion penalty to induce a smoothly-varying tree-like cluster structure, eliminating the need to select the number of clusters \emph{a priori}. We provide an efficient algorithm for convex network clustering and demonstrate its effectiveness on synthetic examples.

\noindent 

\end{abstract}

\begin{figure*}
\centering
\vspace{-0.1in}
\includegraphics[width=\textwidth]{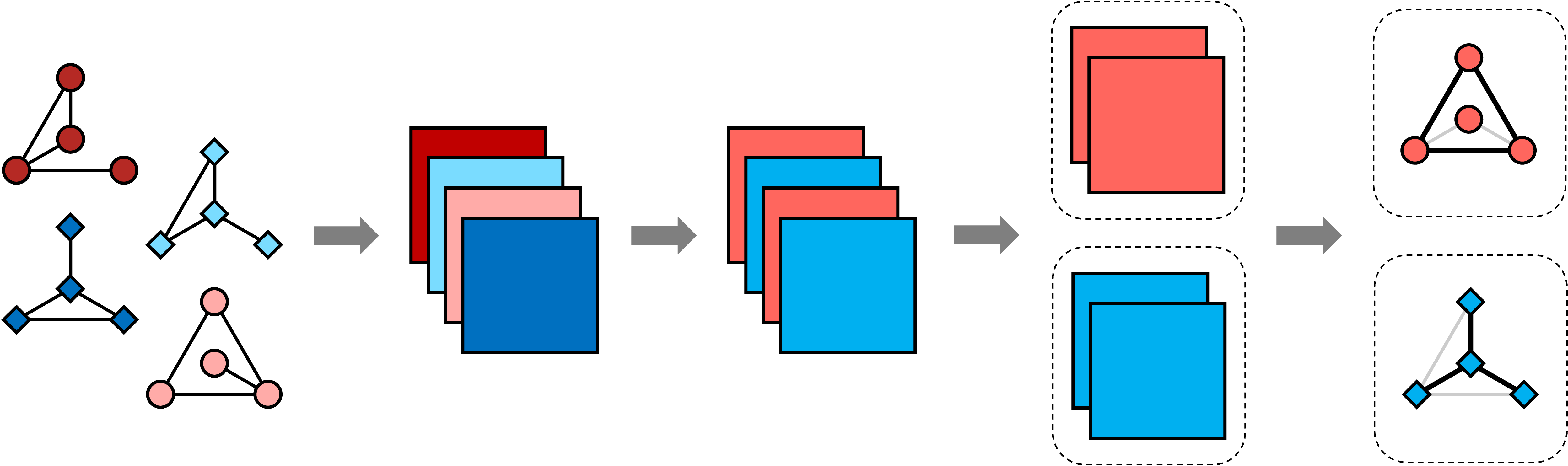}
\caption{A Schematic Representation of Our Approach: Multiple graph observations (first panel) and aligned into a $p \times p \times T$ data tensor, $\Xcal$ (second panel). Application of convex network clustering to the data tensor gives a tensor with replicated slices, $\hat{\Ucal},$ (third panel) which correspond to the centroids of our estimated clusters (fourth panel) and which can be interpreted as ``centroid graphs'' (fifth panel). Note that even though the original observations (first panel) may be unweighted graphs, the estimated centroids (fifth panel) are typically weighted graphs, with edge weights obtained by averaging of all elements of that cluster.}
\label{fig:schematic}
\end{figure*}

\section{Introduction}
The tools of network analysis have proven useful for analyzing a wide variety of data. Networks may be directly observed, \emph{e.g.} social media or telecommunications networks, or may be inferred as statistical models of some underlying phenomenon, \emph{e.g.} genomic or neuroscientific networks \citep{Roddenberry:2021,Yang:2015}. In many situations of interest, it may be possible, or even necessary, to construct multiple network models on the same node set: \emph{e.g.}, a social media network over time or a neuronal firing network under different stimulus conditions. In these contexts, it is often useful to group the networks into scientifically meaningful clusters: this grouping can improve the quality of the estimated network and highlight the common structure present in each group. We propose a novel \emph{convex} method of clustering networks, built on recent convex approaches \citep{Pelckmans:2005,Hocking:2011,Lindsten:2011} to clustering vector data. Our approach inherits the attractive computational and theoretical properties of convex clustering \citep{Chi:2015,Weylandt:2020,Tan:2015,Radchenko:2017,Zhu:2014} and is applicable to almost any class of networks, including directed and undirected, as well as weighted graphs. A case of particular note is a ``network time series'' - that is, networks observed on the same nodes over time.  A minor modification of our approach can be used to induce ``time-structured'' clusters which can be used to detect changepoints or regime-switching dynamics such as networks affiliated with certain latent states. 
\subsection{Background: Convex Clustering}
Convex clustering was originally proposed \citet{Pelckmans:2005} and was later popularized by \citet{Hocking:2011} and by \citet{Lindsten:2011}. This formulation combines a Euclidean (Frobenius) loss function similar to that of $K$-means with a convex fusion penalty reminiscent of hierarchical clustering. The convex clustering solution is given as the solution to the following optimization problem, which clusters the rows of an $n \times p$ data matrix $\bX$:  
\begin{equation}
    \widehat{\bU}=\argmin_{\bU \in \R^{n \times p}} \frac{1}{2} \|\bU - \bX\|_F^2 + \lambda  \sum_{\substack{i, j = 1 \\ i < j}}^n w_{ij} \|\bU_{i \cdot} - \bU_{j\cdot}\|_q. \label{eqn:cvx_clust}
\end{equation}
Here $\lambda \in \R_{\geq 0}$ is a regularization parameter which controls the degree of clustering induced in the matrix of centroids $\widehat{\bU} \in \R^{n \times p}$ and $\{w_{ij}\}$ are non-negative fusion weights which can be used to influence properties of the solution.

Due to the Frobenious loss function, Problem \eqref{eqn:cvx_clust} performs based for Gaussian-like (continuous, symmetric) data. By replacing this loss function, this convex clustering framework has been extended to a variety of other structured data types including: histogram-valued data \citep{Park:2018}, wavelet basis (sparse) data \citep{Weylandt:2021-DSLW}, time series data \citep{Weylandt:2021-ICASSP}, and data drawn from arbitrary exponential families \citep{Wang:2019-GECCO}. The major contribution of this paper is to extend the convex clustering framework to situations where each observation is represented by a network, as discussed in more detail below.

\subsection{Background: Network Clustering}
The task of multiple network clustering has seen relatively little development until recently. \citet{Mukherjee:2017} were among the first to consider the multiple-network problem and propose a spectral-clustering based approach. A different line of approach uses probabilistic models to characterize networks and clusters them accordingly: \citet{Durante:2017} uses a Dirichlet-process (non-parametric Bayesian) prior to cluster networks. \citet{Signorelli:2020} and \citet{Mantziou:2021} extend this type of model-based clustering, with the latter providing an especially thorough review of related work. 

We re-emphasize that our task is grouping together multiple networks on the same node set: not the community detection task of segmenting a single graph which is also sometimes called ``network clustering.'' A convex clustering-based formulation of this latter task was recently developed by \citet{Donnat:2019}. 
\subsection{Contributions}
Our contributions are as follows: we extend the powerful convex clustering framework to multiple network data, yielding a computationally tractable and statistically consistent approach for clustering network observations. Our method takes advantage of the network structure of the data by using a Schatten matrix norm in the fusion penalty, which gives fine-grained control of the intra- and inter-cluster variability. Additionally, we develop an ADMM-type algorithm for the resulting optimization algorithm and establish strong convergence results. Finally, we demonstrate the efficacy of our proposed approach on a variety of synthetic data sets, illustrating its advantages over $K$-means or hierarchical clustering approaches.

\begin{figure*}[t]
\includegraphics[width=0.95\textwidth]{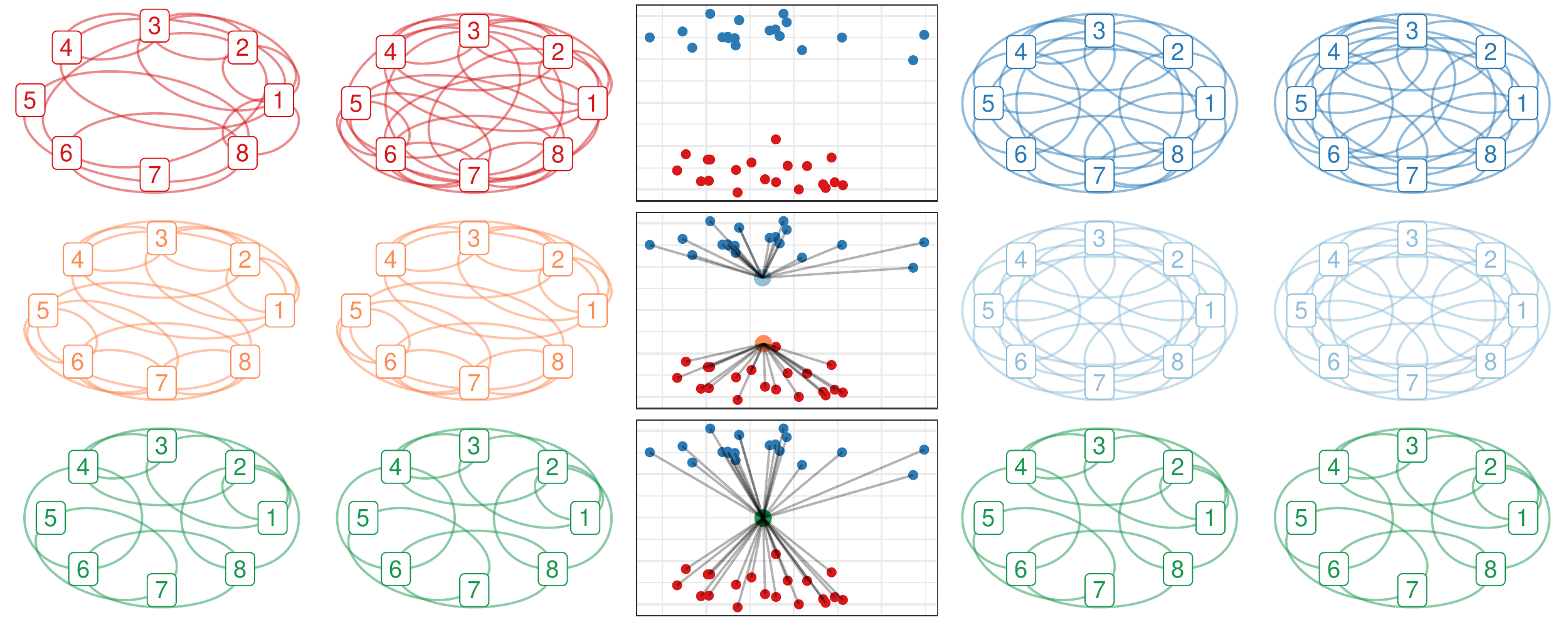}
\caption{Illustration of Convex Network Clustering: Simulated data from a 2-block stochastic block model with different block memberships for two different clusters. Two observations from the first cluster are shown in the left two columns; two observations from the second cluster at the right column. The center column depicts a principal components-type visualization of the networks, computed using the Network PCA model of \citet{Zhang:2019}. At low values of $\lambda$ (top row), observations remain unfused and no clustering is induced. At moderate values of $\lambda$ (middle row), observations in the same cluster are fused together (orange and light blue centroids) but two clusters can be clearly identified. At high values of $\lambda$ (bottom row), all observations are fused to a single cluster (green centroid).}
\label{fig:example}
\end{figure*}

\section{Clustering of Network Series}
We now turn to defining our approach to convex clustering multiple networks: given $T$ networks on $p$ shared nodes $\{\Gcal_t\}_{t = 1}^T$, we construct a $p \times p \times T$ data tensor (multi-dimensional array), $\Xcal$, each slice of which is the adjacency matrix of an observed network. (That is $\Xcal_{\cdot\cdot t}$ is the adjacency matrix of $\Gcal_t$.)
We then solve the following problem:
\begin{equation}
    \hat{\Ucal}^{\lambda} = \argmin_{\Ucal \in \R^{p \times p \times T}} \frac{1}{2}\|\Ucal - \Xcal\|_F^2 + \lambda \sum_{i < j} w_{ij} \|\Ucal_{\cdot\cdot i} - \Ucal_{\cdot\cdot j}\|_{\sigma(q)}
    \label{eqn:network_clustering}
\end{equation}
where $\|\cdot\|_{\sigma(q)}$ denotes the Schatten-$q$ norm, \emph{i.e.}, the $\ell_q$ norm of the singular values of its argument: unless otherwise stated, we take $q = 1$, corresponding to the nuclear norm, in the our experiments, though in some applications $q = \infty$, corresponding to the so-called spectral norm, may be more useful. The use of a matrix norm in the penalty function induces several properties that a vectorized (non-network) approach cannot achieve: specifically, it allows us to control the nature of the differences between cluster centroids. The $q = 1$ nuclear norm induces clusters that differ only by a low-rank matrix while the $q = \infty$ spectral norm encourages centroids with common eigenstructure.

This formulation shares several properties with the standard convex clustering problem \eqref{eqn:cvx_clust}: it combines a Frobenius loss, encouraging the estimated centroids to adhere to the original data, with a ``norm of difference'' penalty function which fuses the estimated centroids together, thereby achieving clustering. For sufficiently large values of $\lambda$, slices of $\hat{\Ucal}^{\lambda}$ are equal: we say that two networks $i$ and $j$ are clustered together if $\hat{\Ucal}^{\lambda}_{\cdot\cdot i} = \hat{\Ucal}^{\lambda}_{\cdot\cdot j}$ and that their common centroid is $\hat{\Ucal}^{\lambda}_{\cdot\cdot i}$.  Figure \ref{fig:schematic} illustrates our approach graphically.

The role of the penalty parameter $\lambda$ is illustrated in Figure \ref{fig:example}: for small values of $\lambda$, no fusions occur and each observation remains in its own cluster located at or near the original data ($\Xcal \approx \hat{\Ucal}$); as $\lambda$ increases, nearby observations are fused together, forming meaningful clusters. As is well known from the penalized regression literature, the penalty term may induce a high degree of bias at this stage and a ``refitting'' step may improve the accuracy of the estimated centroids \cite{Meinshausen:2007,Hastie:2020}; for the clustering problem, this refitting step simply corresponds to the (Frobenius) mean of the cluster members. Finally, at high values of $\lambda$, all observations are fused into a single ``mono-cluster'' whose centroid is the grand mean of all observations. These paths are continuous as a function of $\lambda$ \citep[Proposition 2.1]{Chi:2015} and, in most cases, purely agglomerative \citep[Theorem 1]{Hocking:2011}; hence the clustering path formed as $\lambda$ is varied can also be used to construct a full hierarchical clustering tree \citep{Weylandt:2020,Radchenko:2017}.

We highlight that our approach makes no assumptions on the nature of the graphs $\{\Gcal_t\}_{t = 1}^T$: the graphs can be directed or undirected and our method easily incorporates edge weights. In fact, the graph centroids estimated by Problem \eqref{eqn:network_clustering} are always weighted graphs, even if the data are unweighted. Furthermore, we do not assume any generative model for the graphs, in direct contrast to the model-based clustering literature or methods based on summary statistics.

\subsection{Weight Selection for Structured Clustering}
The choice of fusion weights in Problem \ref{eqn:network_clustering} can significantly change the nature and interpretation of the clustering solution $\hat{\Ucal}^{\lambda}$ \citep{Chi:2019-Trees}. We highlight three weighting schemes of particular importance: 
\begin{itemize}[leftmargin=*]
    \item Uniform or distance-based weights: weights constructed without reference to the graph index $t$ perform traditional clustering. As noted by \citet{Hocking:2011}, adaptive weights based on the inter-observation distances often lead to better performance in practice, but are more difficult to analyze theoretically. In our simulations, we use weights constructed according to the popular truncated RBF scheme \citep{Chi:2015,Chi:2017,Weylandt:2019-ClustRViz}.
    \item Time-based weights: in situations where graphs are observed over time, it is often useful to perform time-based clustering, \emph{i.e.}, changepoint detection. Selecting fusion weights which are non-zero for adjacent observations only will give changepoint type structures. A similar approach was considered for genomic region segmentation by \citet{Nagorski:2018}. 
    \item Hybrid weights: by combining distance- and time-based weights, one can construct weights which encourage clustering patterns which both reflect temporal structure but also have shared centroids at non-consecutive time points. This combination of ``stable + snap-back'' behaviors induces Hidden Markov Model-like dynamics which are particularly useful in the context of latent state detection over time.
\end{itemize}

\begin{figure*}
\centering
\vspace{-0.02in}
\includegraphics[width=\textwidth]{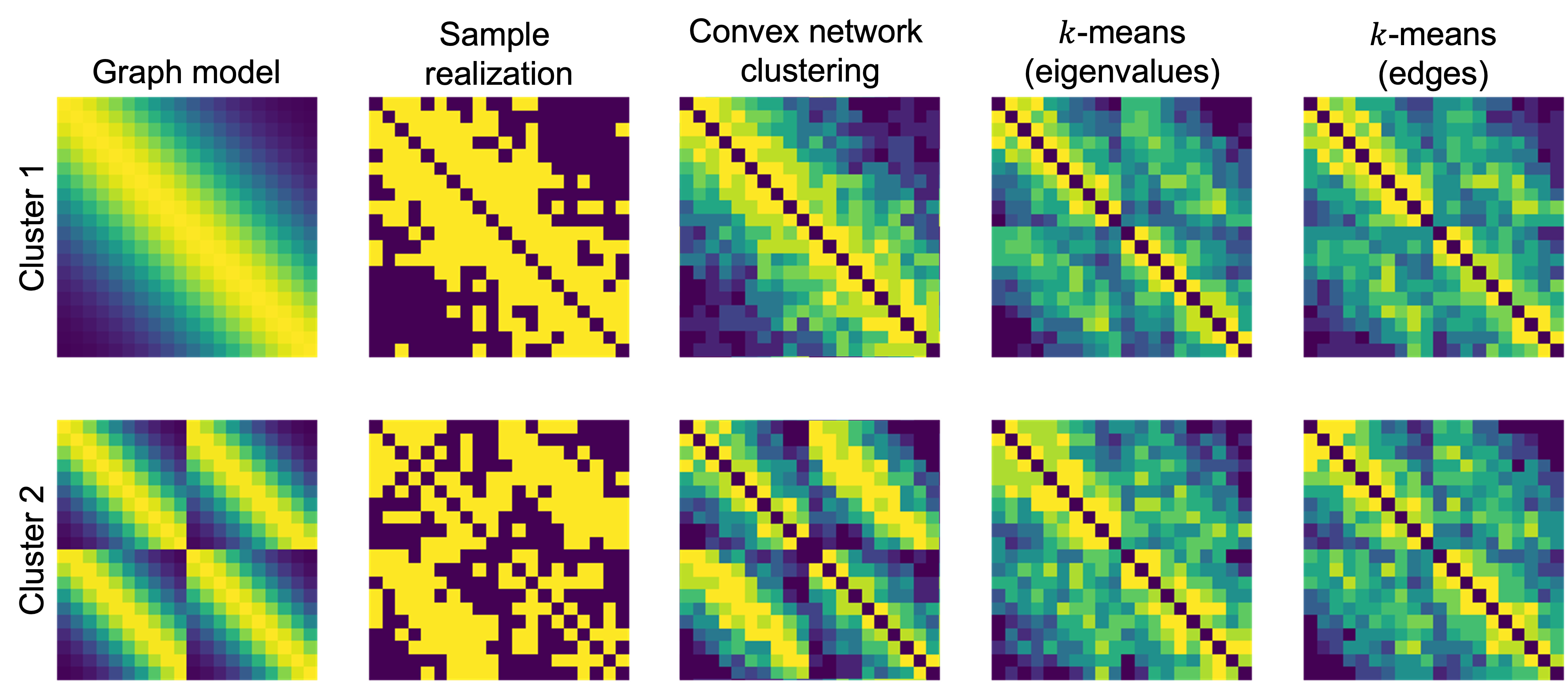}
\caption{Simulation Study from Section \ref{sec:sims}. $T = 20$ graphs on $p = 25$ nodes are simulated from two graph models (top and bottom rows, first column). $K$-means on the eigenvalues (fourth column) or edges (fifth) column only are not able to accurately recover the graph structure while our proposed formulation (third column) accurately recovers the underlying graph models.}
\label{fig:simulation}
\end{figure*}

\section{Computational Considerations}
In addition to the theoretical advantages noted above, Problem \eqref{eqn:network_clustering} is convex and hence admits efficient algorithms for computing the convex clustering solution $\hat{\Ucal}^{\lambda}$. We build on the operator-splitting approach previously considered by \citet{Chi:2015} and by \citet{Weylandt:2019b}, extending to the context of tensor clustering. This yields the following scaled ADMM iterates:
\begin{align*}
    \Ucal^{(k+1)} &= (\mathfrak{I} + \mathfrak{L}^*\mathfrak{L})^{-1}(\Xcal + \rho\mathfrak{L}^*(\Vcal^{(k)} - \Zcal^{(k)}))\\
    \Vcal^{(k+1)} &= \textsf{prox}_{\lambda / \rho P(\cdot; \bw, q)}(\mathfrak{L}(\Ucal^{(k+1)}) + \rho^{-1}\Zcal^{(k+1)}) \\
    \Zcal^{(k+1)} &= \Zcal^{(k)} + \mathfrak{L}(\Ucal^{(k+1)}) - \Vcal^{(k+1)}
\end{align*}
where $\Vcal$ denotes the copy variable, $\Zcal$ the dual variable, $\mathfrak{I}$ the identity operator, $\mathfrak{L}: \R^{p \times p \times T} \to \R^{p \times p \times \binom{T}{2}}$ the operator which calculates all pair-wise differences among the slices of its argument, $\mathfrak{L}^*$ its adjoint, and $P(\cdot; \bw, q)$ the fusion penalty term appearing in Problem \eqref{eqn:network_clustering}, and $\prox_f(\bz) = \argmin_{\bx} f(\bx) + \frac{1}{2}\|\bz - \bx\|_F^2$ denotes the proximal operator. 

While array-oriented programming languages typically implement the addition and scalar multiplication terms appearing in these iterates, it is necessary to express the $\mathfrak{L}$ operator and terms depending on it in a matrix form to put this approach into practice. To do so, we take advantage of the natural isomorphism between $\R^{p \times p \times T}$ and $\R^{p^2 \times T}$, letting $\overline{\cdot}$ denote the action of vectorizing a tensor ``slicewise'' so that each slice of the tensor becomes the row of a $p^2$ column matrix. (In the case of undirected graphs, memory usage can be halved by only using the lower triangle in the isomorphic space, yielding a reduction from $\R^{p^2 \times T}$ to $\R^{\binom{p}{2} \times T}$.) In this space, $\mathfrak{L}$ corresponds to the pairwise difference matrix $\bD$ and $\mathfrak{L}^*$ to its transpose. Our primal iterate thus becomes: 
\begin{align*}
    \overline{\Ucal^{(k+1)}} &= (\bI_{p^2 \times p^2} + \rho\bD^T\bD)^{-1}(\overline{\Xcal} + \rho \bD^T(\overline{\Vcal^{(k)}} - \overline{\Zcal^{(k)}}))
\end{align*}
where $\bI$ denotes the identity matrix. Putting these steps together, we obtain Algorithm \ref{alg:clustering}. By caching the $\mathcal{O}([p^2]^3) = \mathcal{O}(p^6)$ Cholesky decomposition of $\bI + \rho\bD^T\bD$ across iterations, we reduce the complexity of the primal updates to $\mathcal{O}([p^2]^2) = \mathcal{O}(p^4)$ per slice or $\mathcal{O}(p^4T)$ total. The copy update requires taking an SVD of a $p \times p$ matrix for each slice, a $\mathcal{O}(p^3)$ operation, for a total complexity of $\mathcal{O}(p^3T^2)$. Hence the total complexity of each iteratation is $\mathcal{O}(p^4T + p^3T^2)$. Due to the low iteration count typically required by ADMM-type methods, this is sufficient for scaling to problems of moderate size $p \approx 100-1000$, but additional research is required to extend this approach to networks with tens or hundreds of thousands of nodes.

\begin{algorithm}
  \caption{ADMM for Convex Network Clustering}\label{alg:clustering}
  \begin{itemize}
      \item \textbf{Pre-Compute:} 
      \begin{itemize}
          \item Directed Difference Matrix $\bD \in \R^{\binom{T}{2} \times p^2}$
          \item Cholesky factor $\bM = \textsf{chol}(\bI + \rho\bD^T\bD)$
      \end{itemize}
      \item \textbf{Initialize:} $\overline{\Vcal^{(0)}} = \overline{\Zcal^{(0)}}  = \bD\overline{\Xcal}$
      \item \textbf{Repeat Until Convergence:}
          \begin{align*}
    \overline{\Ucal^{(k+1)}} &= \bM^{-T}\bM^{-1}\left[\overline{\Xcal} + \rho \bD^T(\overline{\Vcal^{(k)}} - \overline{\Zcal^{(k)}})\right] \\
    \overline{\Vcal^{(k+1)}}_{\cdot t} &= \prox_{\lambda / \rho\,  w_t\|\cdot\|_{\sigma(q)}}(\bD\overline{\Ucal^{(k+1)}}_{\cdot t} +\overline{\Zcal^{(k)}}_{\cdot t}) & \forall t\\
    \overline{\Zcal^{(k+1)}} &= \overline{\Zcal^{(k)}} + \bD\overline{\Ucal^{(k+1)}}-\overline{\Vcal^{(k+1)}}
\end{align*}
     \item \textbf{Return:} $\Ucal^{(k)}$
  \end{itemize}
\end{algorithm}

We note that Algorithm \ref{alg:clustering} has attractive theoretical properties, including 
primal, dual, and residual convergence for the convex network clustering problem \eqref{eqn:network_clustering}. Furthermore, if $\bD$ has full row-rank, the convergence is linear.  Convergence follows from standard ADMM convergence results, with the linear convergence result being a consequence of the strong convexity of the Frobenius loss and the rank assumptions on $\bD$ \citep{Hong:2017}. Furthermore, Algorithm \ref{alg:clustering} can be incorporated into the \emph{algorithmic regularization} framework of \citet{Weylandt:2020}, allowing the full clustering path, as a function of $\lambda$, to be computed efficiently. We term this approach \textsc{GrassCarp}: \textbf{Gra}ph \textbf{S}pectral \textbf{S}hrinkage \textbf{C}lustering via \textbf{A}lgorithmic \textbf{R}egularization \textbf{P}aths.

\section{Simulation Study} \label{sec:sims}

In this section, we briefly illustrate the superior performance of our method on simulated data. We compare our method to two forms of $k$-means clustering: $k$-means on the spectra, which capture key structural properties of the graphs, and $k$-means on the edge indicator variables, which ignores the network structure. To compare these methods, we simulate $T = 20$ graphs of $p = 25$ notes each from the graphon given by $W(x, y) = 1 - \max(x, y)$ and a equispaced set of sampling points. Realizations from the first cluster are left unchanged while the labels of realizations from the second cluster are permuted to give a distinct network structure. 

Figure \ref{fig:simulation} shows the results of our simulation: $k$-means on the eigenvalues alone does not reflect the permuted labels, leading to poor (essentially random) cluster assignments and inaccurate centroid estimation. By ignoring the spectral properties, $k$-means on the edge struggles with the high-dimensionality and low signal-to-noise ratio of this problem. By contrast, our proposed convex network clustering method is able to take advantage of network spectral structure and edge information and to accurately estimate the cluster assignments and centroids.

\section{Discussion and Extensions}
We have introduced a novel approach for clustering multiple graphs: our approach is based on a convex optimization problem which yields computational tractability and improved statistical performance. We have proposed an efficient ADMM-type algorithm to implement our approach and demonstrate its effectiveness on synthetic data. Our approach combines both edgewise and spectral information via a Frobenius loss and Schatten-norm fusion penalty, yielding a clustering strategy that is robust to a wide range of noise structures, but extensions to more graph-theoretic notions of distance may further improve performance and robustness. While our approach does not make strong assumptions on the observed graphs, it does requires that the same set of label nodes be observed at each observation. This assumption is often violated in practice and extensions of our approach to partially-aligned or unaligned networks are of significant interest. While our method has efficient \emph{per iteration} convergence, the cost of individual iterations is still significant, being dominated by an expensive eigendecomposition for each network at each step. Approximate algorithms that avoid this decomposition are necessary to apply our approach to large-scale networks of popular interest. Finally, we have only considered data-driven clustering formulations: in many tasks, it is reasonable to assume additional properties of the cluster centroids, \emph{e.g.}, low-rank structure, and this information can be incorporated into the clustering procedure. Our method may be useful for clustering multiple networks, detecting changepoints in networks over time, or even for detecting latent network temporal states with applications ranging from analyzing social networks to detecting latent brain connectivity states.

\clearpage
\section{References}
{\ninept \printbibliography[heading=none]}
\end{refsection}

\end{document}